\documentclass[traditabstract]{aa} 
\usepackage{longtable,lscape}
\usepackage{graphicx,color}
\usepackage{txfonts}
\usepackage{natbib}
\usepackage[normalem]{ulem}

\bibpunct{(}{)}{;}{a}{}{,} 

\usepackage{multirow,bigdelim}
{}

\begin{document}
   \title{Explosion of white dwarfs harboring hybrid CONe cores.}
\authorrunning{E. Bravo et al.}
\titlerunning{Hybrid CONe cores explosion}

   \author{E. Bravo
           \inst{1}
   \and
           P. Gil-Pons
           \inst{2}
   \and
           J. L. Guti\'errez
           \inst{2}
   \and
           C.L. Doherty
           \inst{3}
           }

   \institute{Escola T\`ecnica Superior d'Arquitectura del Vall\`es, Univ. Polit\`ecnica de
              Catalunya, Carrer Pere Serra 1-15, 08173 Sant Cugat del Vall\`es, Spain\\   
              \email{eduardo.bravo@upc.edu} 
   \and
	      Departament de F\'\i sica, Univ. Polit\`ecnica de Catalunya, 
Castelldefels, Spain\\
              \email{pilar@fa.upc.edu}; \email{jordi.gutierrez@upc.edu}
   \and
              Monash Centre for Astrophysics (MoCA), School of Physics and Astronomy, Monash University, Victoria 3800, Australia\\
              \email{carolyn.louise.doherty@gmail.com}
	      }

   \date{Received ; accepted }

 
  \abstract{
Recently, it has been found that off-centre carbon burning in a subset of intermediate-mass stars does not propagate all the way to the center, 
resulting in a class of hybrid CONe cores. The implications of a significant presence of carbon in the resulting massive degenerate
cores have not been thoroughly explored so far. Here, we consider the possibility that stars hosting these hybrid CONe cores might 
belong to a close binary system and, eventually, become white dwarfs accreting from a non-degenerate 
companion at rates leading to a supernova explosion. We have computed the hydrodynamical phase of the
explosion of Chandrasekhar-mass white dwarfs harboring hybrid cores, assuming that the
explosion starts at the center, either as a detonation (as may be expected in some
degenerate merging scenarios) or as a deflagration (that afterwards transitions into a delayed
detonation). We assume these hybrid cores are made of a central CO volume, of mass $M_\mathrm{CO}$, 
surrounded by an ONe shell. We show that, in case of a pure detonation, a medium-sized CO-rich region, $M_\mathrm{CO}$ ($ <
0.4$~M$_{{\sun}}$), results in the ejection of a small fraction of the mantle while leaving a
massive bound remnant. Part of this remnant is made of the products of the detonation, Fe-group
nuclei, but they are buried in its inner regions, unless convection is activated during the ensuing
cooling and shrinking phase of the remnant. In contrast, and somehow paradoxically, delayed
detonations do not leave remnants but for the minimum $M_\mathrm{CO}$ we have explored,
$M_\mathrm{CO}=0.2$~M$_{\sun}$, and even in this case the remnant is as small as
0.13~M$_{{\sun}}$.
The ejecta produced by these delayed detonations are characterized by slightly smaller masses of
$^{56}$Ni and substantially smaller kinetic
energies than obtained for a delayed detonation of a ``normal'' CO white dwarf. 
The optical emission expected from these explosions would hardly match the observational
properties of typical Type Ia supernovae, although they make interesting candidates for
the subluminous class of SN2002cx-like or SNIax. 
}


   \maketitle
%

\section{Introduction}

Supernovae are the result of the explosive end-point to the evolution of stars of a wide variety of
masses. Most massive stars go through successive phases of hydrostatic burning until they form an
iron core, at which point they cannot obtain more energy by means of nuclear processes. Their fate
can be either a collapse of the whole structure to form a black hole, or a collapse of the central
region followed by ejection of the envelope. In the last case, the diffusion of radiation
through the ejected material allows it to shine in what is known as a core-collapse supernova
\cite[for recent reviews, see][]{woo05,jan07,jan12,bur13}. In contrast,
stars of smaller mass end as
white dwarfs (WD), essentially dead stars with a composition and mass that is set during their earlier central burning stages, and AGB
phase and associated pulsations. WDs composed of carbon and oxygen (CO) and belonging to
a binary system may be reactivated by accretion of matter from the companion star and, under
favorable conditions, can undergo a final nuclear phase in which they are partially burnt to
Fe-group elements and explode. The nucleosynthesis is rich in unstable isotopes, which decay and
feed the subsequent optical display known as thermonuclear, or Type Ia (SNIa), supernova 
\cite[][]{par07,how11,par14}. The precise path to WD ignition is not completely understood. In
principle, two possibilities are considered, either both components are WDs (the double degenerate
channel, DD), or only one of them is a WD (the single degenerate channel, SD).

There exists an initial mass frontier between core-collapse and white dwarf formation. This frontier is
 located in the region of
intermediate-mass stars (IMS), whose mass lies in the range 
$\sim7$ - 11~M$_{{\sun}}$, with these limits strongly dependent on progenitor metallicity 
and input model physics \cite[][]{woo02,eld04,sma09,ibe13,doh15}. 
Stars in this mass
range go trough a final hydrostatic burning phase transmutating carbon into oxygen and neon (ONe)
\cite[][]{nom84,rit99,sie10}. 
Unlike CO WDs, WDs made of ONe and belonging to a binary system collapse, rather than
explode, when they accrete matter up to the Chandrasekhar-mass, for which reason they are
usually excluded from the set of putative progenitors of SNIa \cite[][]{miy80,nom84,nom87,gut96,gut05}. Furthermore,
merging of two WDs made of CO in a DD scenario may lead to non-explosive C-burning and mutation
of the composition into ONe, with the same result of collapse to a neutron star rather than
explosion \cite[][]{sai98}. Current multidimensional simulations of the merging of WDs nurture the
hope of avoiding the collapse \cite[e.g.][]{pak10,mol14,ras14}. However, these results are
preliminary due to the extremely difficult numerical problems posed by ignition in the outer
shells of the star, usually affected by low spatial resolution in two or three dimensions.  

In this context it is interesting to recall that very efficient neutrino cooling at the innermost 
stellar regions causes 
C-burning start off-centre -- see \cite{dom93} and \cite{gar94} --, except for the most massive cases of the 
above mentioned range.
Eventually C-burning extends further in the core,
but at a certain point the physical conditions required for the burning front to propagate are not met and the process stops, leaving 
a CO central region with a few tenths of M$_{\sun}$ surrounded by an ONe shell.  
Recent analysis \cite[][]{den13} has shown that such a stratified chemical structure of the core might result
due to the effect of convective boundary mixing, which is able to prevent the C-burning flame from 
reaching the central C-rich regions of the star. In fact, similar structures composed of a central C-rich region 
surrounded by an ONe (i.e. C-poor) zone have been found using different evolutionary codes, such as MONSTAR
\cite[][]{doh10}, or the code used by \cite{gar94}. 
The question is nevertheless far from being closed 
\cite[see, e.g.][]{wal07}, 
and it is interesting to analyze the outcome of different configurations of such CONe cores.

The existence of hybrid CONe WD warrants a reevaluation
of the final outcome of the evolution of IMS. The presence of carbon at the center of the WD makes
again feasible the explosion rather than collapse to a neutron star, and the query is whether
the explosion of such a carbon-rich core surrounded by a large, carbon-depleted, ONe mantle
would look similar to a SNIa. 

The width of the initial mass range of IMS that could end as hybrid CONe WD is $\sim0.1-0.2$~M$_{\sun}$ in standard calculations 
\cite[e.g.][]{doh10,ven11,doh15}, but can be as large as $\sim0.4-1.0$~M$_{\sun}$, depending on the strength of convective boundary mixing 
\citep{den13} and uncertainties in C-burning rates \citep{che14}.
We note that CONe cores are found over a large variety of metallicities \citep{doh15} and also for a range of helium-enrichments 
\citep{shi15}.   
In any case, a narrow mass range would allow to explain at most a few percent of the observed SNIa \cite[][]{men14}, small
but important to understand. In particular, it has been advocated that their explosion might
match some class of peculiar SNIa \cite[][]{che14,men14}, and that the narrow initial mass range
would favour the homogeneity of the observational properties within that class.

Currently, there are identified three, more or less homogeneous, classes of sub-luminous
thermonuclear supernovae: the so-called SNIax, the .Ia, and the Ca-rich ones.
The best studied one is the SNIax group \cite[][]{li03,fol13,str15,whi15}, characterized
by maximum-luminosity velocities 2\,000 - 8\,000 km s$^{-1}$ slower and luminosities 0.5 - 5 mag dimmer
than normal SNIa, and hot photospheres. They do follow a different trend in the light curve peak
brightness-width plane, although not far from normal SNIa. They have been tentatively identified
with failed deflagrations of massive WDs leaving a bound remnant. 
Up to now, no Iax supernova has been found in an elliptical galaxy (but SN 2008ge exploded in an S0 galaxy), 
thus suggesting that their progenitors belong to a young population; the galaxies hosting all known Iax supernovae have active regions of
star formation, except for the mentioned case of SN 2008ge. The mass range for stars forming CONe cores is 
consistent with such environments of Iax supernovae. This possibility is further supported by the observed colors of 
the putative progenitors of SN 2008ha and SN 2012Z, 
consistent with those of evolved intermediate mass stars \cite[][]{mcc14b,fol14}. 
The .Ia class \cite[][]{kas10,dro13} name is suggestive of a SNIa with all its relevant
magnitudes scaled down by an order of magnitude: they are characterized by a very fast evolution of
the light-curve, whose peak is about 3 mag dimmer than normal SNIa (i.e., they are about an order
of magnitude less brilliant), and an ejected mass of $^{56}$Ni of order 0.02-0.03~M$_{\sun}$. Their
characteristic velocity, however, is of the same order as normal SNIa, which suggests that the
ejecta mass is small, about a tenth of a Chandrasekhar mass. \citet{dro13} find that the ejecta of
SN2005ek, a .Ia candidate, are dominated by oxygen, while \citet{kas10} speculate that the
progenitor of SN2010X might be an ONeMg WD. Finally, the Ca-rich class
\cite[][]{per10,kas12,val14} is defined by their unusually large content of calcium, that can
amount to as much as half of the ejecta mass. However, at present, their thermonuclear origin has not
been robustly established.

Here, we are concerned with the outcome of the explosion of hybrid WDs, how does their drastic
chemical differentiation influence the hydrodynamic behaviour of the explosion, and which is the
range of properties of their ejecta. We do not calculate the optical outcomes of the
explosions, but concentrate on identifying the most interesting combinations of progenitor system
and explosion type. 
We start by considering the formation of a hybrid CONe WD as a result of the evolution of the primary star in an 
intermediate-mass close-binary system. 
Next, we succinctly explain the methods used in our study and the
characteristics of the initial models. After that, we provide the results of the simulations of the
explosions of hybrid CONe WDs, discuss the results in context with the properties of the
aforementioned subgroups of peculiar SNIa, and conclude.

\section{Formation of hybrid CONe WDs}\label{secform}

The formation of a hybrid CONe degenerate core is a consequence of incomplete core C-burning. 
We consider here two classes of hybrid CONe WDs, characterized by quite different sizes and masses of the CO central region. 
The first class, that we will call here large hybrid-cores (LHC), is obtained following standard evolutionary 
calculations of IMS within a narrow mass-range \citep{doh10,ven11}, and harbor a large CO central volume surrounded 
by a thin ONe layer. They are the result of carbon burning starting very far off--centre in IMS with core masses of 
$\sim1.04-1.06$~M$_{\sun}$, 
and switching-off before reaching the center
due to cooling by neutrinos at the core innermost region. In the cases that concern us, i.e. the lightest objects able to ignite 
carbon, suitable conditions for carbon-burning are not recovered after the first off-center burning episode \cite[for instance][and 
references therein]{rit99,gil01,sie06,doh10}.

The second class, that we will call medium-sized hybrid-cores (MSHC), is the result of IMS evolution when either modified recipes for the 
strenght of convective boundary mixing or uncertainties in the C-burning rates are allowed \citep{den13,che14}. 
The mass of their carbon-rich central volumes is $M_\mathrm{CO}\sim0.2-0.45~\mathrm{M}_{\sun}$, surrounded by an ONe layer thicker than 
that in the LHC class. 
In this class, the quench of the C-burning flame is a result
of the decoupling between the fuel --$^{12}$C-- profile and the thermal profile, 
due to efficient convective boundary mixing \cite[see][]{den13}.
As the flame advances inwards, the carbon abundance it faces decreases
steadily due to the convective mixing of fuel with ashes until, finally, it is so small that carbon ignition is no
longer possible, and a carbon-rich central volume remains unburnt 
\cite[for a recent critical reevaluation of the role of convective 
overshooting see][]{far15}. 

It is interesting to note that in all cases there remains a small 
residual abundance of $^{12}$C in the ONe region, a feature that is in fact common to all Super-AGB star cores.

The scenario that would allow this model become a thermonuclear supernova involves its belonging to a
close binary, whose orbital parameters must meet two requirements. 
First, the initial orbital
period must be such that Roche lobe overflow from the primary, and therefore a mass transfer
episode, will not occur until it has become a giant hosting a deep convective envelope. Second, the
initial mass ratio of the two components, $q=M_1/M_2$ must be relatively low, say below $1.5-2$,
so that merger episodes can be avoided. Both conditions determine initial orbital distances of a
few hundred solar radii and secondary masses $M_2$ above $\approx3-4$~M$_{\sun}$.
Such systems start interacting once the primary component reaches red giant dimensions, and go through a first common envelope (CE)
episode, after which the primary is a He-burning
core and the secondary remains as a main sequence star. 
A second giant (Super-AGB) phase of the primary leads to a new
mass transfer episode, also accompanied by the occurrence of a CE, after
which the primary becomes an almost bare hybrid CONe degenerate core and the secondary still
is a main sequence star. 
The size of the carbon-rich region within the hybrid CONe core
depends on the location of the carbon flame at the onset of the second mass transfer episode, besides other evolutionary ingredients
mentioned above.

Depending on the details of the evolution of orbital parameters, which are beyond the scope of this
work, different options ensue at this point. If the orbital shrinkage has been significant,
gravitational wave radiation might induce further shrinkage and allow Roche-lobe overflow of the
secondary component while it is still a main sequence star. Otherwise, reverse mass transfer can
occur when the secondary reaches a subgiant or giant phase. In either case, if the final outcome is
to be a thermonuclear supernova, instead of a nova, mass accretion rates onto the white dwarf must be above
$10^{-7}~\mathrm{M}_{\sun}/\mathrm{yr}$ \citep{nom07}.


\section{Models and methods}

\begin{figure}[tb]
\centering
  \resizebox{\hsize}{!}{\includegraphics{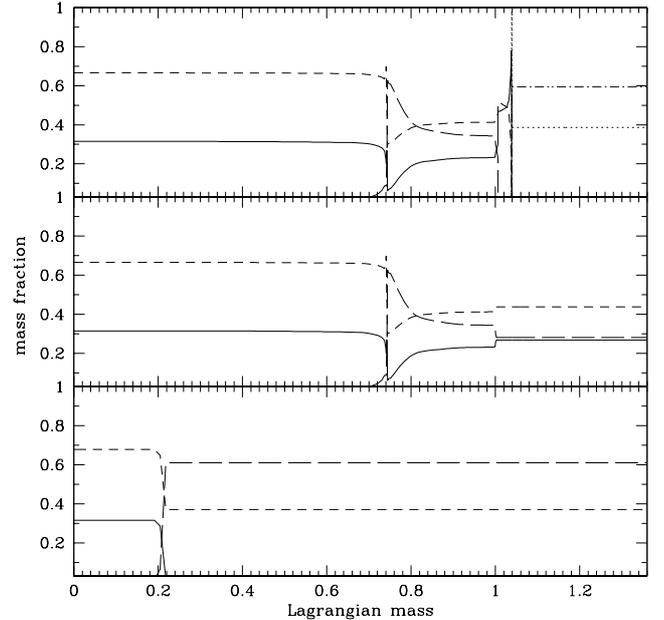}}
\caption{Chemical profile of the AGB core at the beginning of the thermal pulses ({\bf top}) and the pre-explosion WDs of models 22 and 23 
({\bf middle}) and of models 1--6 ({\bf bottom}), see Table~\ref{tab1}. Solid line is C, short-dashed is O, long-dashed is Ne, dotted is 
He, and dot-dashed is H.
}
\label{fig1}
\end{figure}

We have calculated the explosion of both LHC and MSHC Chandrasekhar-mass WDs\footnote{
The explosion of a Chandrasekhar-mass
WD can be representative of the SD as well as the DD channel. In the first case, a hybrid-core WD
accrets matter from a non-degenerate companion until carbon fusion runs away at or near the
center. In the DD case, there are several possibilities, one of them being the merging of both WDs
and the disruption of the less massive to form a corona or accretion disk around the most massive
one. Later, this structure is accreted by the primary at a pace slow enough that there is no
ignition of carbon in its outer regions. In this situation, the evolution of the
accreting WD ends in a similar form as described for the SD scenario.
}
with a version of the
one-dimensional code described in \cite{bra12}, upgraded to include a more accurate treatment of
the coupling between hydrodynamics and nuclear reactions (Bravo et al., in preparation). 
The hydrocode implements a nuclear network of up to 722 nuclides, from free neutrons to $^{101}$In \cite[see Table~I in][]{bra12}. The 
nuclear 
network is solved simultaneously with the hydrodynamic equations and it is adaptive, i.e. its size and membership are 
decided at each time step and at each mass shell according to a set of rules depending on the abundances and reaction rates.

The
initial model fed to the hydrodynamics code was a WD
with central density
$2.6\times10^9$~g~cm$^{-3}$ and total mass $M_\mathrm{WD}=1.37~\mathrm{M}_{\sun}$, in hydrostatic
equilibrium. 
Two
flavors of the explosion of Chandrasekhar-mass WDs were explored, either a direct detonation (DETO)
starting at the center of the star, or a delayed-detonation (DDT) that starts at the center as a
subsonic flame, propagates at a small fraction of the sound velocity (3\% in our models), and later makes a transition to a
detonation at a prescribed density, $\rho_\mathrm{DDT}$. 
The thermal structure of the WD was different for both explosion mechanisms. In the DDT models, we started from an isothermal WD at 
$T=10^7$~K. On the other hand, the initial WD was adiabatic in the DETO models, with a central temperature of $T_\mathrm{c}=10^9$~K. 
Further details of the implementation of the explosion mechanisms in the hydrocode can be found in the Appendix of \cite{bad03}.

In a recent work, \cite{kro15} perform a three-dimensional simulation of a pure deflagration of the carbon-rich core of a hybrid CONe WD, 
assuming that the flame cannot propagate into the ONe mantle. 
The present models complement theirs with respect to the explosion mechanism. We compare results in section~\ref{compkro}.

The chemical composition of the WD and its imprint on the explosion outcome is the main point of interest of this work. To have a realistic 
basis on which to explore different combinations of hybrid core masses, mantle masses, and mass fraction of the main species, we 
have followed the hydrostatic evolution of an IMS star with the aid of MONSTAR \cite[see e.g.][and references therein]{fro96,cam08} in 
the version described in \cite{gil13}. 
The star had a ZAMS mass 8.25~M$_{\sun}$
\footnote{Due to different convective boundary approaches used during the core He burning phase of evolution, we find hybrid CONe cores 
at initial masses $\sim1.5$~M$_{\sun}$ greater than those found in \cite{den13}.}
and solar metallicity, $Z=0.02$ \cite[][]{gre93}, and we followed its evolution up to the
first few thermal pulses during its AGB phase.
In fig.~\ref{fig1} (top panel), we show the composition profiles at the end of our hydrostatic calculation. 
The degenerate
core has a mass of $\sim1.04~\mathrm{M}_{\sun}$, with a structured chemical profile made of a carbon-rich,
$X(^{12}\mathrm{C})\approx0.315$, central volume with $M_\mathrm{CO}\sim0.74~\mathrm{M}_{\sun}$, a thin C-burning front, and a neon-rich 
mantle. 
The transition between the carbon-rich central volume and the carbon-poor mantle is abrupt, a property shared
with the hybrid-cores obtained by \cite{den13}. In
the outermost regions of this core there can be identified thin helium and hydrogen burning
fronts. The whole structure is surrounded by a massive but tenuous hydrogen-rich envelope.
In a binary scenario, relevant for the case we are considering in this work, a first mass-loss
episode will cause the loss of most of the H-rich envelope, and the second mass-loss episode
will remove most of the He-rich remnant envelope. As shown in \cite{gil03}, 
the He and C-burning processes and the resulting core compositions are not altered by 
such mass-transfer, except that the expected progenitor mass would correspond to 0.5 $M_{\odot}$ 
more massive models. Therefore for the present work we can simply use single star evolution models.

In table~\ref{tab1} we summarize the
studied cases. For the MSHC, we have considered carbon-rich regions of masses,
$M_\mathrm{CO}=0.2$, 0.3, and $0.4~\mathrm{M}_{\sun}$ (models 1--19), while for the LHC we have
chosen $M_\mathrm{CO}=0.6$, and $0.74~\mathrm{M}_{\sun}$ (models 20--23). In all
cases, the carbon-rich region is composed of carbon, oxygen and neon, with mass fractions similar to the carbon-rich
region in 
the hydrostatic model shown in
fig.~\ref{fig1}, i.e. $X(^{12}\mathrm{C})=0.315$, 
$X(^{16}\mathrm{O})=0.678$, and $X(^{20}\mathrm{Ne})=0.007$, and surrounded by a carbon-depleted mantle. 

LHC models 22 and 23 mapped
precisely the chemical structure shown in fig.~\ref{fig1} up to 1.0~M$_{\sun}$, to avoid the H and
He burning regions, and then added mass-shells up to $M_\mathrm{WD}$ with the same chemical
composition as in the shell at Lagrangian mass 1.0~M$_{\sun}$
(middle panel in Fig.~\ref{fig1}). 

In all the other cases (models 1-21) the chemical structure of the mantle has been simplified as follows. 
By default, and with the aim of introducing in the explosion models as few free parameters as possible, 
we assume a homogeneous mantle composed of oxygen and neon up to $M_\mathrm{WD}$
(for an example, see the bottom panel in Fig.~\ref{fig1}). 
We have adopted different 
prescriptions for the abundance of oxygen in the mantle, in order to check its impact on the results. 
In most models, we have assumed a mass fraction of $^{20}$Ne in the mantle equal to the maximum one found in fig.~\ref{fig1} (top panel),
$X_\mathrm{out}(\mathrm{^{20}Ne})=0.63$, which belongs to a mass coordinate of $\sim0.74$~M$_{\sun}$, and the 
corresponding $^{16}$O mass fraction, $X_\mathrm{out}(\mathrm{^{16}O})=0.37$.
Since this choice is somewhat arbitrary, we have also explored models with different composition of
the mantle, more reactive than in the default case, either by adding a few percent $^{12}$C
(models 2 and 8), or by increasing $X_\mathrm{out}(\mathrm{^{16}O})$ at the expense of $X_\mathrm{out}(\mathrm{^{20}Ne})$ (see
column five in table~\ref{tab1}, and section~\ref{composec}). Finally, we have explored the presence of a mantle formed by
two layers, an inner one with the default composition mentioned above, and an external one rich
in carbon, $X_\mathrm{ext} (^{12}\mathrm{C})=0.32$,with a mass $M_\mathrm{ext}$ (column six). In section~\ref{thesize} we show 
that 
the precise structure of the mantle has no impact on the outcome of the explosion of these models.

\newcommand\multibrace[3]{\rdelim\}{#1}{3mm}[\pbox{#2}{#3}]}
\begin{table*}[tb]
\caption{Summary of results of the simulated explosions of hybrid-core CONe WDs.}\label{tab1}
\centering
\begin{tabular}{rlccccclllllclll} 
\hline\hline  
ID &           
$M_\mathrm{CO}$ & Burning & $\rho_\mathrm{DDT,7}$\tablefootmark{b} &
$X_\mathrm{out}(\mathrm{^{16}O})$ &
$M_\mathrm{ext}$ & $M_\mathrm{rem}$ &
$M(^{56}\mathrm{Ni})$\tablefootmark{c} & $M(\mathrm{IME})$ & $M(\mathrm{IGE})$ & $K$\tablefootmark{d} & Comment \\
& (M$_{\sun}$) & mode\tablefootmark{a} & & (M$_{\sun}$) & & (M$_{\sun}$) &
(M$_{\sun}$) & (M$_{\sun}$) & (M$_{\sun}$) & (foes) & \\
\hline
1 & 0.2 & DETO & - & 0.37 & - &  1.30 & 0 & - & - & & no optical display &
\rdelim\}{19}{6pt}{} & \multirow{19}{*}{\rotatebox[origin=c]{90}{MSHC}} \rule{0pt}{2.6ex} \\
2 & 0.2 & DETO & - & 0.37\tablefootmark{e} & - &  1.30 & 0 & - & - & & no optical display \\
3 & 0.2 & DETO & - & 0.37 & 0.57  & 1.28 & 0 & 0.002 & - & & no optical display \\
4 & 0.2 & DDT & 1.1 & 0.37 & - &  0.12 & 0.08 & 0.021 & 0.095 & 0.06 & slow ejecta \\
5 & 0.2 & DDT & 1.4 & 0.37 & - &  0.13 & 0.07 & 0.023 & 0.072 & 0.05 & slow ejecta \\
6 & 0.2 & DDT & 1.4 & 0.37 & 0.57  & 0.014 & 0.10 & 0.085 & 0.18 & 0.09 & slow ejecta \rule[-1.2ex]{0pt}{0pt}
\\
7 & 0.3 & DETO & - & 0.37 & - &  1.00 & 0 & - & - & & no optical display \\
8 & 0.3 & DETO & - & 0.37\tablefootmark{e} & - &  0.88 & 0 & - & - & & no optical display \\
9 & 0.3 & DETO & - & 0.63 & -  & 1.00 & 0 & - & - & & no optical display \\
10 & 0.3 & DDT & 0.9 & 0.37 & -  & - & 0.27 & 0.489 & 0.374 & 0.52 & SNIax? \\
11 & 0.3 & DDT & 1.1 & 0.37 & -  & - & 0.35 & 0.490 & 0.461 & 0.63 & SNIax? \\
12 & 0.3 & DDT & 1.4 & 0.37 & -  & - & 0.44 & 0.450 & 0.583 & 0.73 & SNIax? \\
13 & 0.3 & DDT & 1.7 & 0.37 & -  & - & 0.54 & 0.395 & 0.681 & 0.80 & SNIax? \\
14 & 0.3 & DDT & 2.0 & 0.37 & -  & - & 0.63 & 0.355 & 0.759 & 0.85 & SNIax? \rule[-1.2ex]{0pt}{0pt} \\
15 & 0.4 & DETO & - & 0.37 & -  & - & 0.10 & $<10^{-3}$ & 0.414 & 0.09 & slow ejecta \\
16 & 0.4 & DETO & - & 0.63 & -  & - & 0.95 & 0.004 & 1.36 & 1.28 & almost no IME \\
17 & 0.4 & DETO & - & 0.37 & 0.57  & - & 0.10 & $<10^{-3}$ & 0.414 & 0.09 & slow ejecta \\
18 & 0.4 & DETO & - & 0.37 & 0.77  & - & 0.10 & $<10^{-3}$ & 0.415 & 0.09 & slow ejecta \\
19 & 0.4 & DDT & 1.4 & 0.37 & - &  - & 0.46 & 0.437 & 0.597 & 0.76 & SNIax? \rule[-1.2ex]{0pt}{0pt} \\
20 & 0.6 & DETO & - & 0.37 & - &  - & 0.95 & 0.004 & 1.36 & 1.30 & almost no IME &
\rdelim\}{4}{6pt}{} & \multirow{4}{*}{\rotatebox[origin=c]{90}{LHC}} \\
21 & 0.6 & DDT & 1.4 & 0.37 & -  & - & 0.52 & 0.386 & 0.649 & 0.82 & SNIax? \rule[-1.2ex]{0pt}{0pt} \\
22 & 0.74 & DETO & - & 0.37 & -  & - & 0.95 & 0.003 & 1.36 & 1.43 & almost no IME \\
23 & 0.74 & DDT & 1.4 & 0.37 & -  & - & 0.59 & 0.451 & 0.710 & 1.09 & SNIax? \rule[-1.2ex]{0pt}{0pt} \\
\multicolumn{12}{l}{\underbar{Models with mixed initial composition\tablefootmark{f}}} \\
24 & 0.2--0.4\tablefootmark{g} & DETO & - &  & - &  - & 0.95 & 0.003 & 1.36 & 1.29 & almost no IME \\
25 & 0.2--0.4\tablefootmark{g} & DDT & 1.4 &  & - &  1.26 & 0 & - & - & & no optical display \\
26 & 0.2--0.4\tablefootmark{h} & DETO & - &  & - &  - & 0.95 & 0.002 & 1.36 & 1.37 & almost no IME \\
27 & 0.2--0.4\tablefootmark{h} & DDT & 1.4 &  & - &  - & 0.59 & 0.427 & 0.712 & 0.99 & SNIax? \\
28 & 0.74\tablefootmark{i} & DDT & 1.4 &  & - &  - & 0.61 & 0.443 & 0.726 & 1.11 & SNIax? \rule[-1.2ex]{0pt}{0pt} \\
\multicolumn{12}{l}{\underbar{Reference models without a hybrid core}} \\
29 & 1.37 & DDT & 1.1 & 0.5 & - &  - & 0.54 & 0.491 & 0.697 & 1.29 & normal SNIa \\
30 & 1.37 & DDT & 1.4 & 0.5 & - &  - & 0.66 & 0.409 & 0.818 & 1.35 & normal SNIa \\
31 & 1.37 & DDT & 1.7 & 0.5 & - &  - & 0.74 & 0.347 & 0.904 & 1.40 & normal SNIa \\
32 & 1.37 & DDT & 2.0 & 0.5 & - &  - & 0.79 & 0.299 & 0.964 & 1.42 & normal SNIa \\
\hline
\end{tabular}
\tablefoot{
\tablefoottext{a}{Burning mode: DETO = pure detonation; DDT = delayed detonation.}
\tablefoottext{b}{Density of deflagration-to-detonation transition in DDT models, in $10^7~\mathrm{g}~\mathrm{cm}^{-3}$.}
\tablefoottext{c}{This column and the next two give the ejected mass of \element[][56]{Ni}, intermediate-mass elements (IME), and 
iron-group elements (IGE), respectively.}
\tablefoottext{d}{Kinetic energy of the ejecta (1 foe = $10^{51}~\mathrm{erg}$).}
\tablefoottext{e}{Initial abundance of $X(^{12}\mathrm{C})$ set to 0.02 outside $M_\mathrm{CO}$.}
\tablefoottext{f}{WD born with hybrid core, but mixed during carbon simmering.}
\tablefoottext{g}{Model with $X(^{12}\mathrm{C})=0.10$, representative of the result of the mixing between hybrid cores with 
$M_\mathrm{CO}$ within the range given in column two and a mantle with its default composition, 
$X_\mathrm{out}(\mathrm{^{16}O})=0.37$ and 
$X_\mathrm{out}(\mathrm{^{20}Ne})=0.63$.}
\tablefoottext{h}{Model with $X(^{12}\mathrm{C})=0.20$, representative of the result of the mixing between hybrid cores with 
$M_\mathrm{CO}$ within the range given in column two and a mantle formed by
two layers, an inner one with the default composition and an external one rich
in carbon, $X_\mathrm{ext} (^{12}\mathrm{C})=0.32$, with a mass $M_\mathrm{ext}=0.57$~M$_{\sun}$.}
\tablefoottext{i}{Initial composition obtained by full mixing of that of the initial model of ID 22 (and ID 23).}
}
\end{table*}

The initial models described above assume that the chemical differentiation established during the
He-burning and quenched C-burning Super-AGB phase is maintained up to the supernova explosion in the form of a hybrid-core. 
However, in the last stages of evolution prior to thermal
runaway of a Chandrasekhar-mass WD, the energy released by the incipient carbon fusion reactions
is enough to drive convection \cite[][]{pir08b,pir08}\footnote{Another mechanism of chemical
mixing would be gravitational settling, whose efficiency would depend on the time elapsed since
the formation of the hybrid-core WD and the supernova event.}. The convective core that develops
is usually thought to encompass most of the WD, which in our case would destroy the chemically
differentiated structure of the hybrid core. Nevertheless, there are uncertainties about the
feasibility of such a huge
convective core. First, in general, the eventual activation of URCA pairs might limit the reach of
convective motions \cite[][]{ste06,pod08,den15} and preserve the original chemical structure.
Second, specific to hybrid-core WDs, the presence of an abrupt separation between the
carbon-rich central volume and the carbon-depleted mantle might allow for the operation of a
self-regulating mechanism restricting the extent of convection. 
Convection during carbon simmering is regulated by the release of nuclear energy near the center of the 
WD. When convection reaches the ONe mantle and mixes it with the material in 
the core, the carbon mass fraction in the center is expected to decrease 
(besides the effect of carbon consumption by nuclear reactions). As a result, 
the nuclear energy generation rate may decrease, which would affect the reach of convective 
motions. Through this mechanism, the carbon-depleted
mantle may provide a natural frontier for the convective core, avoiding full mixing with the
carbon-rich central volume.
However, a proper determination of the extent of such effect, even if it is relevant at 
all, would require following the evolution of the hybrid WD during the carbon 
simmering phase, which is beyond our scope.

In view of the above uncertainties, full mixing of the WD prior to explosion cannot be discarded.
To explore its consequences, we have calculated additional models of the explosion of homogeneous WDs 
representative of the full mixing of some of the hybrid-core WDs discussed before.
(models 24--28 in table~\ref{tab1}).
We address these homogeneous models in section~\ref{simmsec}.

Finally, models 29--32 are reference delayed-detonation models with varying $\rho_\mathrm{DDT}$, in
which the initial WD had a homogeneous composition, i.e. without a differentiated core, of $X(^{12}\mathrm{C}) = X(^{16}\mathrm{O}) \simeq 0.5$,
with traces of heavier species with solar metallicity ratios. These models, that do not come from hybrid-core WDs, 
are identified in Table~\ref{tab1} with $M_\mathrm{CO} = 1.37$~M$_{\sun}$. This kind
of models have been successfully compared to Type Ia supernovae and remnants in the past. 

\section{Results}

\begin{figure}[tb]
\centering
  \resizebox{\hsize}{!}{\includegraphics{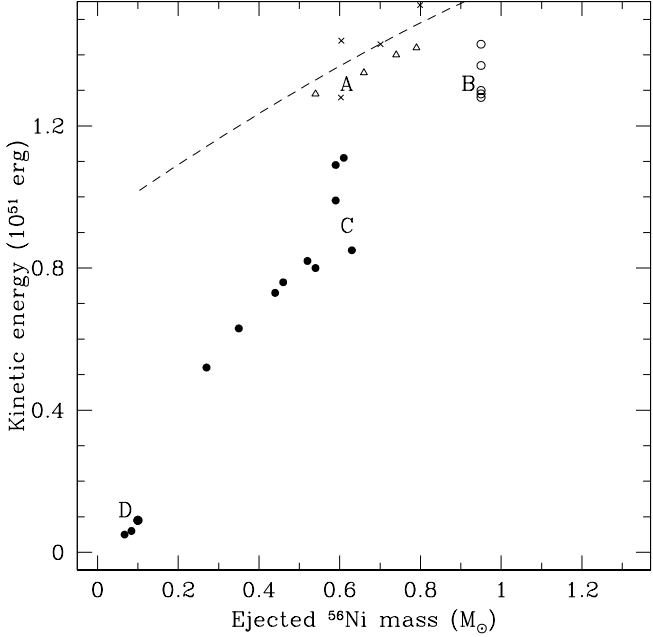}}
\caption{Location of the computed explosions of hybrid CONe WDs in the
$K$-$M(^{56}\mathrm{Ni})$
plane, in context with standard SNIa models. Empty circles
stand for pure detonations of hybrid WDs, while solid circles are for delayed detonations of the
same objects. Empty triangles belong to delayed detonations of homogeneous CO WDs with varying
$\rho_\mathrm{DDT}$, which can account for normal SNIa (models 29--32 in Table~\ref{tab1}). 
DETO models 15, 17, and 18 cannot be distinguished because their location coincides with that of DDT model 6.
Crosses belong to the models of 
\cite{ohl14}, while the dashed line shows the approximate
location of the normal SNIa models according to \cite{maz07} (from .08~M$_{\sun}$ of $^{56}$Ni for the faint SN1991bg to .94~M$_{\sun}$ for 
the bright SN1994ae) and \cite{woo07} (who provide a range of kinetic energies). 
The four regions labelled with letters  
group models with more or less homogeneous characteristics (see text for further details).
}
\label{fig2}
\end{figure}

The results of our supernova simulations are shown in table~\ref{tab1} and in
fig.~\ref{fig2}. We start by discussing the results of the simulations of our default configuration,
i.e. homogeneous mantle depleted of carbon and with oxygen abundance $X_\mathrm{out}(^{16}\mathrm{O})=0.37$.
The dependence on several parameters of the initial model is discussed later.

\subsection{Explosion of large-hybrid-core (LHC) Chandrasekhar-mass WDs}

The outcome of the explosion of LHC models with $M_\mathrm{CO}$ of 0.6 and 0.74~M$_{\sun}$ is quite
similar, the main difference deriving from the lower nuclear binding energy of the composition
of the WD with the smaller $M_\mathrm{CO}$. These explosions completely disrupt the WD leaving no bound
remnant. Both DETO models (20 and 22) incinerate almost the whole WD and eject a huge mass of
$^{56}$Ni, but almost no intermediate-mass elements (IME). Since the clear detection of IME at
maximum light is one of the defining properties of SNIa, these pure detonation models cannot match
observations, just as with the detonation of homogeneous CO WDs. 

On the other hand, DDT models
(21 and 23) produce amounts of $^{56}$Ni and IME in agreement with the predictions of normal
SNIa models, but the kinetic energy of the ejecta is, respectively, 0.82 and 1.09 foes,
substantially smaller than those attributed to typical SNIa for their $M(^{56}\mathrm{Ni})$ \cite[see, e.g.,][and 
fig.~\ref{fig2}]{maz07,woo07}.
In both DDT models, there is ejected a small mass of
unburned $^{16}$O, 0.14~M$_{\sun}$, while the mass of $^{12}$C is negligible.
The similarity of the results of $M_\mathrm{CO}=0.6$ and 0.74~M$_{\sun}$ shows that the composition of the mantle is not a 
relevant factor determining the outcome of the explosion of LHC models.

\subsection{Explosion of medium-sized hybrid-core (MSHC) Chandrasekhar-mass WDs}

The result of the explosion of WDs with $M_\mathrm{CO}$ in the range $0.2 - 0.4~\mathrm{M}_{\sun}$,
requires more explanation. Pure detonation models 1, 7, and 15, are characterized by a small
release of nuclear binding energy. In all of them, the detonation consumes the whole carbon-rich
region, but is unable to propagate into the mantle due to the reduced flammability of
oxygen as compared to carbon. However, the result of the explosion depends sensitively on 
$M_\mathrm{CO}$, because we are considering masses close to the minimum incinerated mass necessary
to unbind a Chandrasekhar-mass WD. The $M_\mathrm{CO} = 0.2~\mathrm{M}_{\sun}$ case releases too low nuclear
energy and the explosion
leaves a large remnant of $1.30~\mathrm{M}_{\sun}$, i.e. only $0.07~\mathrm{M}_{\sun}$ are ejected
composed of oxygen and neon. The outcome of the detonation of the $M_\mathrm{CO} = 0.3~\mathrm{M}_{\sun}$ model is
similar, with the only difference that the ejected mass is larger, $0.37~\mathrm{M}_{\sun}$, still
composed exclusively of oxygen and neon. No significant optical display, similar to any flavour of
supernova, is expected from such objects. Nevertheless, the bound structures would be inflated and
hot for some time, which would make them brighter than inert WDs of the same mass and potentially 
detectable at not too large distances.

The pure detonation model with $M_\mathrm{CO} = 0.4~\mathrm{M}_{\sun}$ is energetic enough to unbind the
whole WD, even though just a very small fraction of the mantle is burnt. The final kinetic energy, 0.09 foes, 
as well the ejected mass of $^{56}$Ni, $0.10~\mathrm{M}_{\sun}$, are quite small.

The result of the DDT models varies qualitatively as a function of $M_\mathrm{CO}$
but, in all cases studied, more mass is burnt than in the corresponding DETO models. For 
$M_\mathrm{CO} = 0.2~\mathrm{M}_{\sun}$ (models 4 and 5), there remains a small bound remnant of
$\sim0.1$~M$_{\sun}$, while the ejecta has a small kinetic energy, a few hundredths of
$10^{51}$~erg, and small $^{56}$Ni mass. These results are quite like the ones obtained for the pure
detonation of the $M_\mathrm{CO} = 0.4~\mathrm{M}_{\sun}$ case, and do not depend significantly on the value of
$\rho_\mathrm{DDT}$. 

The result of the delayed-detonation changes dramatically for $M_\mathrm{CO}>0.2~\mathrm{M}_{\sun}$. 
Models 10--14 and 19 are similar to 
the DDT
models of LHC, i.e. the WD is completely disrupted and the amount of $^{56}$Ni ejected is within
that expected for normal SNIa, while the kinetic energy is relatively small. There is a monotonic
increase of both the mass of $^{56}$Ni and the kinetic energy as a function of $M_\mathrm{CO}$, for
a given $\rho_\mathrm{DDT}$. For instance, at $\rho_\mathrm{DDT}=1.4\times10^7$~g~cm$^{-3}$, the
range of $^{56}$Ni masses covered by DDT models of both LHC and MSHC leaving no bound remnant is
0.44--0.59~M$_{\sun}$, and the range of kinetic energies 0.73--1.09~foes. 
Allowing for different
DDT transition densities, the amount of $^{56}$Ni synthesized monotonically increases with $\rho_\mathrm{DDT}$, similar to that
found for the reference models 29--32,
but the kinetic energy remains much smaller than in the
latter. 

\begin{figure}[tb]
\centering
  \resizebox{\hsize}{!}{\includegraphics{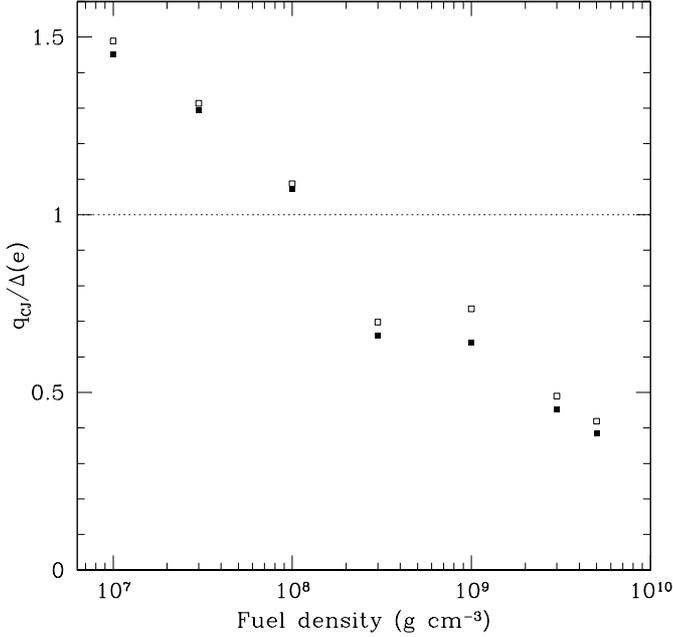}}
\caption{Ratio of the nuclear energy released in a Chapman-Jouguet detonation, $q_\mathrm{CJ}$, to
the increase in specific internal energy of the fuel needed to achieve the shocked 
state, $\Delta(e)$, as a function of fuel density. Open squares belong to a fuel composition of
$X(^{12}\mathrm{C}) = X(^{16}\mathrm{O}) = 0.5$, while solid squares are for initial 
composition $X(^{12}\mathrm{C}) = 0.2$ and $X(^{16}\mathrm{O}) = 0.8$. A dotted line indicates the minimum $q_\mathrm{CJ}$ necessary 
for self-sustained detonation propagation $q_\mathrm{CJ} = \Delta(e)$.
}
\label{fig3}
\end{figure}

It is remarkable that, in all the MSHC models just discussed, the most violent burning mode, the pure detonation, leads to explosions less
energetic than the delayed-detonation. This
apparent paradox can be understood as resulting from pure detonations entering the ONe
mantle at high densities, combined with a monotonically decreasing nuclear energy release in 
detonations as a function of fuel density. Since pure detonations do not allow for any hydrodynamic
response of the surrounding matter, the densities at which they arrive at the ONe mantle are
the same as in the initial structure at carbon runaway, that is above $10^9$~g~cm$^{-3}$ for 
$M_\mathrm{CO} = 0.2-0.4~\mathrm{M}_{\sun}$. Figure~\ref{fig3} shows the ratio of the nuclear energy
released in a Chapman-Jouguet detonation to the increase in specific internal energy of the fuel
within the shock front (just before nuclear reactions commence), as a function of fuel density. At
fuel densities below $\sim10^8$~g~cm$^{-3}$, with little dependence on the precise composition of
the fuel, the energy released within the reaction layer of a detonation is enough to drive the fuel
to the shocked state, where it starts burning and releasing energy that allows more fuel to
detonate \cite[][]{kho88}. In these conditions, the detonation can self-sustain once it has been
initiated. In contrast, at higher densities, the final state of the detonation ashes is so hot that
a substantial fraction of the produced iron-group elements (IGE) photodisintegrate and the final
products are a mixture of IGE, IME, and light elements, with the corresponding reduction in the
nuclear energy yield. The nuclear energy invested in the creation of IME and light
elements is restored later, due to their recombination when matter expands and density decreases by about one order of magnitude or
more, but this cannot occur while the detonation is alive because of its supersonic nature.
Thus, the nuclear energy released at large densities is insufficient to shock the fuel to the desired state, and 
a piston would be needed\footnote{Usually, such a piston-like effect is provided in pure detonation models
by the supersonic burning of a region close to the center of the WD, as a result of a shallow
thermal gradient, which gives a phase velocity of the combustion front above the sound speed.}, or
another source of energy external to the detonation itself, to sustain the burning wave. As a
result, the pure detonation dies shortly after entering the ONe mantle. 

In the DDT models, burning propagation is initially subsonic, as a deflagration, leaving
time for matter to expand before arriving to the ONe mantle. The densities at which the
detonation front arrives to the mantle, e.g. in models 12 and 19, are of order $10^7$~g~cm$^{-3}$, low
enough to allow self-sustainability of the burning front. The exception are the models with 
$M_\mathrm{CO} = 0.2~\mathrm{M}_{\sun}$, e.g. model 5, in which the flame reaches the mantle before
the detonation initiation has succeeded, and finally fails to burn the WD. 

\subsection{Parametric dependencies}

We examine here what are the variations, if any, introduced by changing the initial setup of the
explosion models.

\subsubsection{Composition of the mantle}\label{composec}

Models 9 and 16 are similar to models 7 and 15 (pure detonations of $M_\mathrm{CO} = 0.3$ and
$0.4~\mathrm{M}_{\sun}$), respectively, but with a higher $^{16}$O abundance in the ONe mantle.
The change of the composition of the mantle has no effect on the explosion of the model with
the smaller $M_\mathrm{CO}$. However, the detonation of the $M_\mathrm{CO} = 0.4~\mathrm{M}_{\sun}$ case is
completely different. With the larger $^{16}$O abundance, the piston formed by the detonated
carbon-rich region is energetic enough to allow the incineration of almost the whole WD. This shows
how a moderate change in the initial conditions can have a huge impact on the final outcome of the
explosion of these structures, and that the $M_\mathrm{CO} = 0.4~\mathrm{M}_{\sun}$ case is on the frontier between those that leave a
massive bound-remnant and those that produce a ejecta very rich in $^{56}$Ni. 

We explored, in models 2 and 8, the effect that a small residual abundance of
$^{12}$C in the ONe mantle has on DETO models with $M_\mathrm{CO} = 0.2$ and $0.3~\mathrm{M}_{\sun}$ cases (to be compared
to models 1 and 7). For the smaller $M_\mathrm{CO}$, there is no change at all. For the
$0.3~\mathrm{M}_{\sun}$ case the only change is that the detonation penetrates slightly more in the
ONe mantle, not enough to make a healthy explosion, but the ejecta mass is larger,
$0.49~\mathrm{M}_{\sun}$ instead of $0.37~\mathrm{M}_{\sun}$. As the ejecta is in any case devoid
of radioactive isotopes, this change is not expected to have any observational consequences. 

\subsubsection{Structure of the accreted mantle}\label{thesize}

More complex structures of the mantle have been explored in models 3, 6, 17, and 18, which have
a thick carbon-rich layer surrounding the ONe shell. The results, which have to be compared
to models 1, 5, and 15, show no significant impact of the presence of such chemically
differentiated mantle, but for slight differences in the remnant mass in the case of a DDT explosion.

\subsubsection{Mixing during carbon simmering}\label{simmsec}

Convective mixing during carbon simmering is a potential source of homogeneization of the WD composition prior
to explosion. Assuming full chemical mixing of the initial WD models with 
$M_\mathrm{CO} = 0.2$, 0.3, and $0.4~\mathrm{M}_{\sun}$ and default mantle composition, i.e. without carbon, the global $^{12}$C 
mass fractions are $X(^{12}\mathrm{C})=0.05$, 0.07, and 0.09, respectively. 
Contrary to the results of chemically differentiated hybrid-cores described in previous sections, the composition and structure of the 
mantle may have an influence on the outcome of fully mixed models. This effect is particularly relevant if the mantle is composed by two 
layers, one of them rich 
in carbon, as assumed in some of the models in Table~\ref{tab1} and section~\ref{thesize}. For instance, assuming full chemical mixing of 
the initial WD models with 
$M_\mathrm{CO} = 0.2$, 0.3, and $0.4~\mathrm{M}_{\sun}$, a middle layer with no carbon and an external layer of 
$M_\mathrm{ext}=0.57$~M$_{\sun}$ with $X_\mathrm{ext} (^{12}\mathrm{C})=0.32$, the global $^{12}$C 
mass fractions are $X(^{12}\mathrm{C})=0.18$, 0.20, and 0.23, respectively. 
We have computed the explosion (DETO and DDT modes) of two models representative of the class 
of homogenized MSHC models, one in which $X(^{12}\mathrm{C})=0.10$ and the other with $X(^{12}\mathrm{C})=0.20$ (models 24 -- 27 in 
Table~\ref{tab1}).

Pure detonation of these homogeneous models (24 and 26) are similar to the successful\footnote{
Here, by successful we mean that the detonation progressed through a 
substantial fraction of the WD, as in models 16, 20, and 22, whereas in models 15, 17, and 18 half of the WD mantle remained with its 
original composition.
}
pure detonations
of hybrid-cores with $M_\mathrm{CO}$ from 0.4 to $0.74~\mathrm{M}_{\sun}$. They are able to propagate
across most of the WD, unbinding the whole star and ejecting a huge mass of $^{56}$Ni. The impact
of mixing on the result of pure detonations is, thus, to eliminate the diversity on the outcomes,
in particular the possibility that the explosion leaves a massive bound remnant. 

The delayed detonation of the homogeneous model with $X(^{12}\mathrm{C})=0.10$ (model 25), 
is not able to unbind the WD completely. The small mass fraction of $^{12}$C makes the
release of nuclear energy too small to initiate a healthy detonation, and the burning
front dies after having consumed $\sim0.35$~M$_{\sun}$. The final result is rather similar
to that of the pure detonation of the hybrid-core case with $M_\mathrm{CO} = 0.2~\mathrm{M}_{\sun}$.

The delayed detonation of the homogeneous model with $X(^{12}\mathrm{C})=0.20$ (model 27), leads to outcomes similar to those obtained from 
the hybrid core cases with the same $\rho_\mathrm{DDT}$ and $M_\mathrm{CO} > 0.2~\mathrm{M}_{\sun}$. 

We have also computed the explosion (DDT model 28) of a homogeneous model with chemical composition obtained from the full mixing of the 
hybrid core model 23 (middle panel in Fig.~\ref{fig1}). In this case, the global carbon mass fraction is $X(^{12}\mathrm{C})=0.28$, quite 
close to its value in the hybrid core of model 23. Indeed, the outcome of the delayed detonation of both models reported in 
Table~\ref{tab1} is within 2--3\% of each 
other. All in all, successful delayed-detonations of homogeneous
models seem to encompass a narrower range of ejecta properties than their hybrid-core counterparts.

We conclude that the main effect of the eventual destruction of the chemically differentiated
structure (hybrid-core) during the carbon simmering phase is a reduction on the variation of the
possible outcomes, but they do not affect qualitatively the properties of successful explosions. 

\subsubsection{Comparison to \cite{kro15}}\label{compkro}

The initial model in \cite{kro15} was a cold chemically-differentiated WD with central density $2.9\times10^9$~g~cm$^{-3}$. Their 
chemical structure consisted of a carbon rich, $X(^{12}\mathrm{C})=0.5$, core of $M_\mathrm{CO}=0.2~\mathrm{M}_{\sun}$ surrounded by an 
ONe layer of 0.9~M$_{\sun}$ (with a 3\% carbon abundance by mass) and topped by a CO layer of 0.3~M$_{\sun}$. This initial model is most 
similar to those of our models 1-6. 

The main difference between our calculations and those of \cite{kro15} lies on the assumed geometrical configuration of the flame at 
runaway, a major unknown in theoretical studies of SNIa explosion. While our simulations start, necessarily (due to the one-dimensional 
nature of our code), from a spherically symmetric volume at the center of the WD, they ignite five kernels off-center. These kernels soon 
merge, forming a one-sided plume that floats towards the surface while the flame grows in size. Finally, in the simulation of \cite{kro15} 
less than half the CO core is consumed by the deflagration, and only 0.014~M$_{\sun}$ of material is ejected, of which 
$3.4\times10^{-3}~\mathrm{M}_{\sun}$ are $^{56}$Ni. 

Unlike the model of \cite{kro15}, in all of our explosion models 1-6 the whole CO core is consumed by the burning front. Their final 
results are intermediate between those of our models 1-3 and 4-6. In the first group, there is left a massive remnant, as 
in \cite{kro15} model, but no $^{56}$Ni is ejected. In the second group, there is a small amount of $^{56}$Ni ejected, but at a quite low 
velocity due to the large mass of the ejecta. We speculate that, since the initial geometry of the flame may not be always the same, it is 
possible that all of these models are realized in nature.

\section{Discussion}

\begin{figure*}[tb]
\centering
  \resizebox{\hsize}{!}{\includegraphics{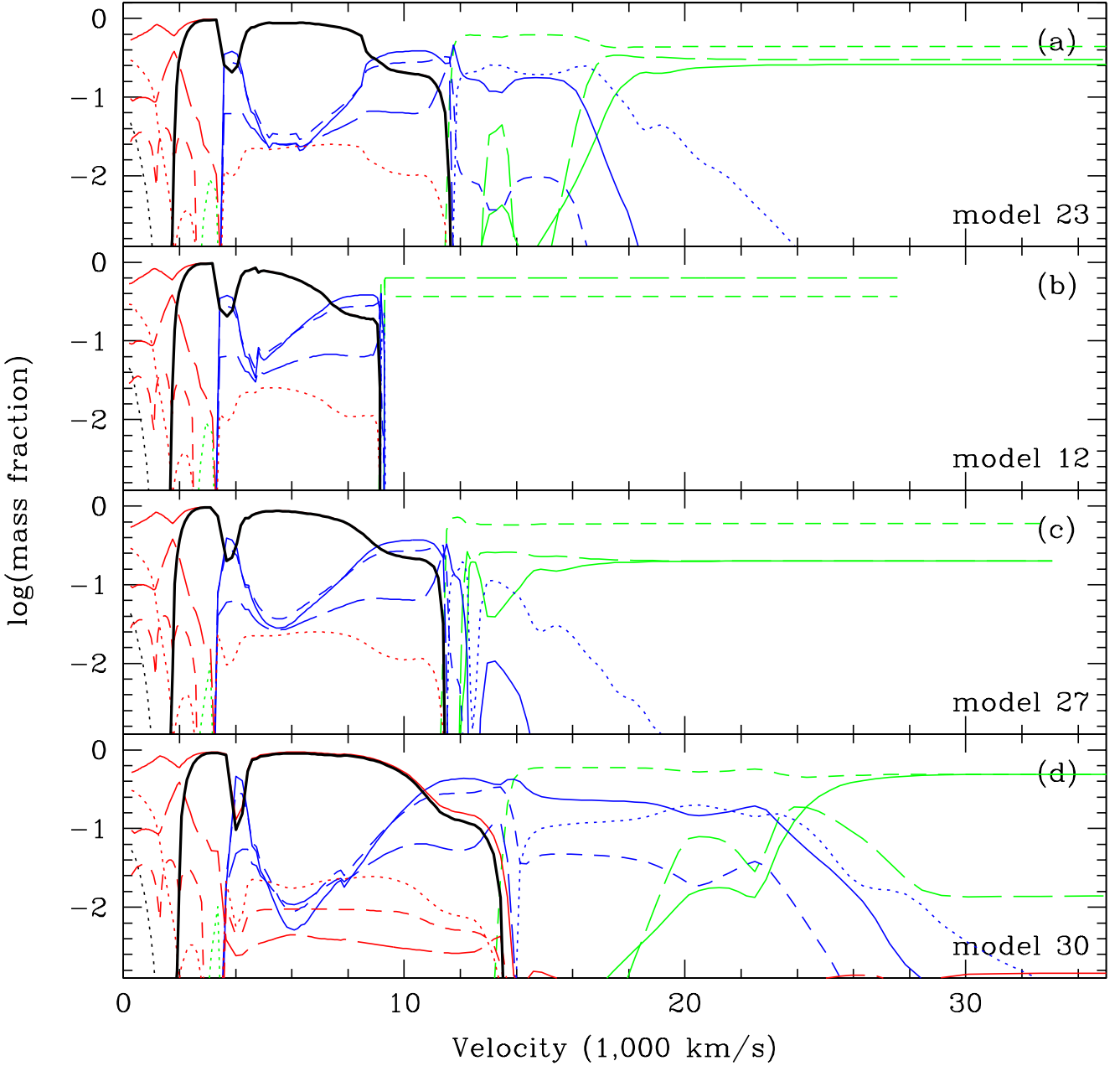}}
\caption{
Chemical composition of the ejected matter in four delayed detonation
models with $\rho_\mathrm{DDT}=1.4\times10^7$~g~cm$^{-3}$. From top to bottom, hybrid WD with a $M_\mathrm{CO} = 
0.74$~M$_{\sun}$ (model 23), hybrid WD with $M_\mathrm{CO} = 0.3$~M$_{\sun}$ (model 12), mixed model with initial $X(^{12}\mathrm{C})=0.20$ 
(model 27), and pure CO WD with an initial homogeneous composition 
with $X(^{12}\mathrm{C}) = X(^{16}\mathrm{O}) \simeq 0.5$ and solar mass fractions of species with $A\ge24$ (model 30). The black
thick line is the abundance by mass of $^{56}$Ni a few seconds after explosion. The rest of lines
give the abundance by mass of the most important elements after radioactive decays (i.e., the final
stable isotopes), and have the following meaning: green stands for fuel (solid is C, short-dashed
is O, long-dashed is Ne); blue stands for IME (solid is Si, dotted is Mg, short-dashed is S,
long-dashed is Ca); and red stands for IGE (solid is Fe, dotted is Cr, short-dashed is Mn,
long-dashed is Ni). 
}
\label{fig4}
\end{figure*}

The ultimate verification of a supernova model relies on the comparison of synthetic spectra and
light curves with observations. However, our purpose in this work is to identify the most promising
models for future computation of their optical properties. To this aim, we base the following
discussion on the location of the computed models in the kinetic energy vs ejected $^{56}$Ni mass
plane (fig.~\ref{fig2}) and in the chemical profile as a function of velocity (fig.~\ref{fig4}).

In fig.~\ref{fig2} we plot our reference models 29--32 (open triangles) together with the models computed by \cite{ohl14}, who explored the 
delayed detonation of Chandrasekhar-mass WDs with homogeneous
composition and variable abundance of $^{12}\mathrm{C}$, 0.2--0.5 in mass fraction (crosses). These models
(labelled ''A'' in the plot) help identify the combination of kinetic energy and
ejected mass of $^{56}$Ni necessary to obtain a normal SNIa, $K\simeq1.0-1.5\times10^{51}$~erg and
0.3--0.8~M$_{\sun}$ of $^{56}$Ni \cite[][]{maz07,woo07}. The range of $^{56}$Ni masses can be extended down to 0.08~$M_{\sun}$ and up to 
0.94~M$_{\sun}$ to account for the full set of SNIa that follow the light-curve luminosity-width relation (dashed line) \cite[][]{maz07}.

Pure detonation models that burn the ONe mantle (models 16, 20, 22, 24, and
26 in table~\ref{tab1}, or ``B`` in the plot) lie to the right of the normal SNIa models, with
similar kinetic energy but much larger $M(^{56}\mathrm{Ni})$. These models burn almost all of the
WD at high density so they do synthesize at most a few thousandths of a solar mass of IME, which is
in clear contradiction with the requirement of prominent Si and S absorption features that
define SNIa (aside from the absence of hydrogen). Such a kind of object has never been observed.
On the other hand, pure detonations that did not succeed in burning a significant fraction of the
ONe mantle leave a huge remnant and eject from a few hundredths to a few tenths of a solar mass
of material with the original composition of the mantle, i.e. with no radioactive isotopes. These
explosions would not be detectable as a supernova, and are labelled as ''no optical display`` in
table~\ref{tab1}. Their remnants would expand due to its large energy content and the radioactive input, and would remain as
brighter-than-normal ONe WDs during a Kelvin-Helmholtz timescale. 

Region ''C`` in fig.~\ref{fig2} is populated by delayed detonations of hybrid-core WD with
carbon-rich central volumes of 0.3~M$_{\sun}$ or more, either chemically differentiated 
(models 10--14, 19, 21, 23) or homogeneous (models 27 and 28) at
thermal runaway. This group of models is characterized by slightly smaller amounts of $^{56}$Ni
ejected and substantially smaller kinetic energies than normal SNIa models. Both tendencies point
towards the subluminous class of SNIax. \cite{whi15} distinguishes two different populations
within the SNIax class. One, that they call SN2002es-like, characterized by small dispersion of
light-curve properties with rise times slightly faster than normal SNIa, slow ejecta, and a
preference for early-type host galaxies. Their ejecta mass is thought to be of order
$\sim0.5~\mathrm{M}_{\sun}$. The other, that they call after SN2002cx, presents a larger dispersion
of light-curve shapes but the rise times are consistently slower than in normal SNIa, the ejecta
speed is slightly faster than in the SN2002es-like population, and they are usually hosted by
late-type galaxies. Their ejecta mass can be as large as the Chandrasekhar-mass,
depending on the effective opacity assumed by \cite{whi15} \cite[but see][]{mcc14}. 

The velocity distribution of the explosion products in some delayed detonation models located in region ''C``
is shown in fig.~\ref{fig4}, together with the result of the reference delayed detonation model 30. 
Panel (a) displays the velocity distribution for the LHC DDT model 23. 
Comparing with the bottom panel, (d), it is clear that
the LHC models are distinguished by Si and S extending up to much lower velocities than in the
reference model\footnote{In this comparison, we disregard the effects due to the inclusion of solar
metallicity abundances in the pre-explosive structure of the reference model, which leaves an
imprint in the $^{56}$Ni-rich core in the form of large Mn and Ni abundances, and in the outermost
layers as the presence of iron.}. For instance, $X(\mathrm{Si})\ge0.1$
below $\sim24,000$~km~s$^{-1}$ in the reference model, while the limit is
$\sim16,000$~km~s$^{-1}$ in the uppermost panel, i.e. $\sim8,000$~km~s$^{-1}$ slower. On the
other side, the tip of the $^{56}$Ni-rich core is $\sim2,000$~km~s$^{-1}$ slower in the hybrid-core
model than in the reference model. The relatively short distance between the radioactive core and
the Si-rich region might explain the hotter than normal photospheres deduced for the SNIax class.

Panels (b) and (c) of fig.~\ref{fig4} display the chemical structure of delayed detonations of both chemically
differentiated and homogeneous models coming from $M_\mathrm{CO} = 0.3$~M$_{\sun}$, MSHC models
12 and 27. The structure of the $^{56}$Ni-rich inner ejecta is quite similar
to the other models shown in the same plot, still Si and S extend to much smaller velocities, so they are
even closer to the radioactive core than they are in the LHC models. This trend is suggestive of a
correlation between dimmer objects and hotter photospheres, but such a correlation has not yet been reported for SN2002cx-like objects.

The SNIax group extend to quite low luminosities that require less than a hundredth of a solar mass
of $^{56}$Ni ejected, e.g. SN2008ha or SN2007qd. Models labelled as ''slow ejecta'' in
table~\ref{tab1} (models 4--6, 15, 17, and 18, ``D`` in fig.~\ref{fig2}\footnote{
Note that models 15, 17, and 18 cannot be distinguished in this fig. because their location coincides with that of DDT model 6.}
) produce and eject similar
amounts of $^{56}$Ni. They are able, as well, to account for the small specific kinetic energy of
the ejecta required to explain these supernovae. However, the ejecta mass in the models is close to
the Chandrasekhar mass, while these events are usually explained as low-mass ejecta
\cite[][]{mcc14,whi15}. Since the energy released in these models is close to the bounding energy
of the progenitor WD, we cannot discard that some multidimensional hydrodynamics effect can
redistribute the kinetic energy in such a way that a larger remnant is left.
Interestingly, \cite{kro15} show that an asymmetrical pure deflagration of a hybrid-core Chandrasekhar-mass WD 
can match the observational properties of SN2008ha, one of the faintest SNIax ever recorded,
although the ejected mass seems to be too small to account for this particular supernova. Nevertheless, the model was 
not specially tuned to reproduce the observables of SN2008ha, and the parameter space was not explored to find a better agreement. So, it 
is 
likely that relatively small differences in the initial model can result in a better fit of the properties of SN2008ha.

With respect to the other aforementioned subclasses of subluminous SNIa, both the .Ia and the
Ca-rich ones are not compatible with any of the models we have calculated. The .Ia class is
characterized by a very fast rise of the light curve, which is at odds with the large mass of the
ejecta we obtain (unless there operates some multidimensional hydrodynamics effect, as mentioned
in the previous paragraph). On the other side, we do not see any model that is particularly rich in
calcium, as observed in the Ca-rich group.

\section{Conclusions}

We have contributed to the study of the impact of hybrid-core WDs on
the understanding of SNIa and their diversity. We
have calculated the explosion of Chandrasekhar-mass hybrid-core WDs, with different masses of the
carbon-rich central volume. We aim to identify the combination of evolutionary scenario and explosion
mechanism that provide potential models for peculiar SNIa, and whose light curve and spectra should
be addressed in future works.

Pure detonations of Chandrasekhar-mass CO WDs have been dismissed for decades as possible
mechanisms of SNIa explosion, because they turn almost the whole star into IGE ashes. However,
the presence of an ONe mantle in hybrid-core WDs opened the possibility that pure detonations
succeed in producing enough intermediate-mass elements to match the observational properties of, at least, some subgroup
of SNIa. Our results show that this is not the case. Indeed, the presence of an ONe
mantle may have a strong impact on the propagation of a detonation, but the outcome is quite
different from any kind of thermonuclear supernova observed so far. If the carbon-rich region is too
small, $M_\mathrm{CO}\lesssim0.3~\mathrm{M}_{\sun}$, the detonation dies shortly after entering
the ONe mantle, with the result that the nuclear energy release is too small and most of the WD
remains bound after the thermonuclear event. A few tenths of a solar mass can be ejected, but
devoid of radioactive elements. As a result, nothing like a supernova is to be expected. On
the other hand, if the carbon-rich region is large enough that the propagation of the detonation
through the ONe mantle can be sustained by the piston action of the incinerated core, then almost
the whole WD is burned to ashes. In such a case, as in the pure CO composition case, the ejecta would contain
no IMEs, in contradiction with SNIa observations. The same result is obtained if the chemical
composition is mixed just before thermal runaway, during the carbon simmering phase, even for
small masses of the carbon-rich central volume. We conclude that pure detonations in hybrid-core WDs can be
discarded as the explosion mechanism behind any known class of SNIa, just as for pure CO WDs. 

But, if the explosion mechanism is of the delayed detonation kind, the properties of the ejecta from hybrid-core WDs are reminiscent of
the subluminous class SNIax, in particular the SN2002cx-like population \cite[][]{whi15}. They are
characterized by slower velocities than our reference models, that can explain normal SNIa, and a
smaller extent of the Si and S-rich layers, which are closer to the $^{56}$Ni-rich core. We suggest
that the proximity of the IMEs to the radioactive sources may explain the hot photospheres often
detected in SNIax, which would imply an inverse correlation between photosphere temperature and
supernova luminosity. Our simulations of delayed detonations of hybrid-core WDs are scarcely
sensitive to the mixing of the chemical composition during the carbon simmering phase. Another appealing feature of hybrid-core WDs as SNIax 
progenitors is that they come from stars more massive than CO WDs do
and the secondary must be heavier than $3-4~\mathrm{M}_{\sun}$ (see section~\ref{secform}),
which may explain in a natural way the preference of SN2002cx-like
objects for galaxies with active stellar formation regions. 

On the other hand, our models do not cover the whole range of $^{56}$Ni ejected masses that
seems to be implied by current observations of SNIax. 
Specifically, we do not find models that
leave a massive remnant and eject a $^{56}$Ni mass as small as 0.1~M$_{\sun}$ or less (the smallest amount
of $^{56}$Ni in our models is 0.27~M$_{\sun}$, but its precise value depends on the assumed
$\rho_\mathrm{DDT}$).
This could be related to the limitations implicit in our one-dimensional
study of the explosion of hybrid-core WDs. For instance, our delayed-detonation explosions of
carbon-rich central volumes of 0.2~M$_{\sun}$ eject a few hundredths of a solar mass of $^{56}$Ni, but the whole
ejecta mass is too large for the imparted kinetic energy.
In this case, the expected light curve would be much wider than observed.
In an asymmetrical event, some bullets of
$^{56}$Ni-rich matter could float through the ONe mantle during the deflagration phase of the
explosion, redistributing the energy in such a way that the ejecta mass were smaller
\cite[][]{bra06,bra09a}, i.e. closer to that needed to explain the dimmest SNIax. 

The explosions we have calculated are restricted to Chandrasekhar-mass WDs, that could arise either
from the SD or the DD (slow merger) scenario. However, hybrid-core WDs may experience a
thermonuclear explosion in other scenarios, like double-detonations of subChandrasekhar-mass WDs,
explosions of Chandrasekhar-mass WDs surrounded by a low-density halo (coming from a DD merger),
violent mergers, and violent collisions, several of which require multidimensional simulations.
We will explore such kind of problems in future works. 

\begin{acknowledgements}
We thank the referee for useful suggestions that have improved the presentation. This  work  was supported by Spanish MINECO grants 
AYA2012-33938 and AYA2013-40545.
\end{acknowledgements}

\bibliographystyle{aa}
\bibliography{../ebg}

\end{document}